# Storm fronts over galaxy discs: Models of how waves generate extraplanar gas and its anomalous kinematics

Curtis Struck[1*] and Daniel C. Smith[2*]

[1] Dept. of Physics & Astronomy, 12 Physics Bldg., Iowa State University, Ames, IA 50011, USA
[2] Applied Physics Lab., 1100 Johns Hopkins Rd., Johns Hopkins University, Laurel, MD, 20723, USA
[*] E-mail: curt@iastate.edu (CS), Daniel.Smith@jhuapl.edu (DCS)

## ABSTRACT

The existence of partially ionized, diffuse gas and dust clouds at kiloparsec scale distances above the central planes of edge-on, galaxy discs was an unexpected discovery about 20 yrs ago. Subsequent observations showed that this EDIG (extended or extraplanar diffuse interstellar gas) has rotation velocities approximately 10-20% lower than those in the central plane, and have been hard to account for. Here we present results of hydrodynamic models, with radiative cooling and heating from star formation. We find that in models with star formation generated stochastically across the disc an extraplanar gas layer is generated as long as the star formation is sufficiently strong. However, this gas rotates at nearly the same speed as the mid-plane gas. We then studied a range of models with imposed spiral or bar waves in the disc. EDIG layers were also generated in these models, but primarily over the wave regions, not over the entire disc. Because of this partial coverage, the EDIG clouds move radially, as well as vertically, with the result that observed kinematic anomalies are reproduced. The implication is that the kinematic anomalies are the result of three-dimensional motions when the cylindrical symmetry of the disc is broken. Thus, the kinematic anomalies are the result of bars or strong waves, and more face-on galaxies with such waves should have an asymmetric EDIG component. The models also indicate that the EDIG can contain a significant fraction of cool gas, and that some star formation can be triggered at considerable heights above the disc midplane. We expect all of these effects to be more prominent in young, forming discs, to play a role in rapidly smoothing disc asymmetries, and in working to self-regulate disc structure.

**Key words:** galaxies: ISM, galaxies: structure, Galaxy: disc

## 1 INTRODUCTION

It is natural to expect that the typical vertical extent above the disc mid-plane of the warm phases of the interstellar gas (T ≈ 3000-8000K) in galaxies like the Milky Way is roughly equal to its thermal pressure scale height (see e.g., Spitzer 1978). Nearly two decades ago observational studies of edge-on disc galaxies began to reveal counter-examples, such as the well-known NGC 891 disc (e.g., Dettmar 1990; Rand, Kulkarni & Hester 1990). In recent years observations in the radio continuum (e.g., Dahlem, Lisenfeld & Rossa 2006), hydrogen alpha line (Rossa & Dettmar 2003, Rossa, Dettmar & Walterbos 2004, Heald et al. 2006a,b, 2007), dust emission (Howk 2005), mid-infrared (Burgdorf, Ashby & Williams 2007), X-ray bands (Tuellmann et al. 2006b, Wang 2007), and the radio 21 cm line (Barbieri 2005, Oosterloo, Fraternali & Sancisi 2007, Rand & Benjamin 2008) have provided much more information on extended gas in this and other edge-on systems. It now seems clear that the phenomenon of warm gas at distances of more than several times the thermal scale height above disc mid-planes is not rare, especially given limited observational detection sensitivities. The Milky Way also possesses a comparably extended layer of diffuse ionized gas, the Reynolds Layer (e.g., Reynolds 1990). Gas far from the galactic mid-plane is detected by a variety of means, and seems to include the "intermediate velocity cloud" population (e.g., Kerton, Knee & Schaeffer 2006).)

The evidence suggests that the gas mass in the layer is correlated with high global rates of star formation (SF), see Dahlem, Lisenfeld & Rossa (2006), and Tuellmann et al. (2006a). It should be noted that this phenomenon extends over large radial ranges in the disc, and in contrast to galactic winds or superwinds, is not the result of central starbursts or nuclear activity in galaxies. Moreover, there is no evidence for overall outflow of this extended diffuse interstellar gas (henceforth EDIG); it is distinct from the galactic wind phenomenon.

Rossa, Dahlem, Dettmar, & van der Marel (2008) also find that the morphology of the EDIG (e.g., its filamentary structure) depends on the level of SF activity. Similarly, Li, et al. (2008) find specific extraplanar, diffuse X-ray features associated with disc SF sites in their high resolution observations of NGC 5775. Thus, the EDIG-SF activity association may be local as well as global. The energy sources associated with young star activity, including stellar winds, supernovae, and high UV and cosmic ray fluxes, are certainly sufficient to support the EDIG.

The discovery that the EDIG rotates significantly less rapidly than gas in the central plane at comparable galactic radii also provided an important clue to its nature. Azimuthal velocity gradients of order tens of km/s/kpc have been observed in the vertical direction (Swaters, Sancisi & van der Hulst 1997, Heald et al. 2006a,b, 2007, Oosterloo, Fraternali & Sancisi 2007, Fraternali et al. 2007).

Several theoretical explanations have been proposed to account for the existence of the EDIG and its kinematic peculiarities. Foremost of these is the model of multiple galactic fountains or chimneys, which postulates that gas clouds are launched from the vicinity of young star clusters, and once away from those sites the clouds follow ballistic trajectories above or below the disc. This idea, which dates from the 1970s (Shapiro & Field 1976), implicitly assumes that these regions are empty of obstacles like other clouds. The observations apparently contradict this, and recent studies show that the ballistic fountain model fails to reproduce the EDIG kinematics (Benjamin 2002, Fraternali & Binney 2006). The failure of this seemingly self-evident model is another surprising result.

Alternate theories include infall from accreted satellites (Fraternali & Binney 2006). Fraternali has recently provided several good candidate accretion cases (Fraternali 2008). However, infall models have to be rather fine-tuned to account for the observed symmetries and trends in EDIG properties across the different systems that have now been observed.

Barnabé et al. (2006) have described models of baroclinic (but not barotropic) gas distributions, which have a steady structure and reproduce the kinematic anomalies. Although nonthermal energy sources, such as turbulence and shocks might produce such flows, this has not been explicitly demonstrated. (See Struck & Smith (1999) for an example of a related, turbulently supported disc model, which is appropriate for discs with high rates of SF).

The purpose of the present work is to present exploratory numerical hydrodynamical models to try to gain further insight into the problem, and provide direction for future research. The models described in this paper fit in a region of a space of models between those produced for a variety of related, but different purposes. For example, there is a significant literature of models on the development of superbubbles or fountain flows (e.g., the recent work of Joung & Mac Low 2006, Tenorio-Tagle, Muñoz-Tuñón, Silich, & Telles 2006, Tenorio-Tagle, Wünsch, Silich & Palous 2007, and Melioli, Brighenti, D'Ercole, & de Gouveia Dal Pino 2008). There is also a growing literature on high resolution modeling of the turbulent structure of the interstellar medium (e.g., de Avillez & Breitschwerdt 2005, 2007, Kritsuk, Norman, Padoan, & Wagner 2007, Wada & Norman 2007, and Hennebelle, et al. 2008). In both these applications the modelled region usually has a scale size of order a kpc, though in some cases adaptive meshes allow resolution on much smaller scales, and a much larger scale-length may be used in one direction (e.g., perpendicular to the disc). As will be discussed below we focus on disc-wide phenomena, like waves, which we believe play a more important role in producing the EDIG than isolated fountains. Thus, we do not study the latter, and do not have sufficient resolution for comparison to the best models of interstellar turbulence.

In recent years several groups have modelled the structure and phase balance in whole discs, and investigated the role of different cooling and feedback formalisms (e.g., Springel & Hernquist 2003, Tasker & Bryan 2006, 2008, Booth & Theuns 2007, Booth, and Theuns, & Okamoto 2007, Robertson & Kratsov 2008). As in our own unpublished work (Smith 2001), the general conclusion is that global structure and phase balance do not depend sensitively on these prescriptions, though structural details do (e.g., Tasker & Bryan 2008).

In line with our goal of using exploratory models to learn about the EDIG, our models were designed to be relatively simple compared to some of these works. A Smooth Particle Hydrodynamics algorithm, with a large number of particles, was used to model the disc gas. The particles were initially set on circular orbits, bound by the gravity of a rigid halo potential. In addition, a fixed gravitational potential, representing the effects of a stellar disc, was applied in the direction perpendicular to the disc plane (the z-direction, see below). Gas pressure was computed with an adiabatic equation of state, and optically thin radiative cooling was applied to warm/hot particles. Energy input from young star activity was applied for a finite duration to particles exceeding a threshold density, and having temperatures below a set maximum. This input generated turbulent pressure, though the effects of other nonthermal pressure sources (e.g., magnetic fields, cosmic rays, and photo-pressure on dust grains) were not included.

Further details on the simulation code are given in the following section. The results of model runs for three different cases – flocculent discs, discs with imposed spirals, and discs in a barred potential – are given in Section 3. Models with two different surface densities (called LD (low density) and HD (high density)) were run for each of the three cases. The results are summarized and prospects for further research given in Section 4.

## 2 SIMULATION CODE

### 2.1 General Features of the Simulation Code

In this work we have attempted to keep the simulation code as simple as possible, consistent with the primary goal of modelling a few important dynamical processes in the EDIG. The reasons for this approach include making the code reasonably fast, so that a number of simulations could be run to explore a range of initial conditions with modest computer resources. A second reason was to make the code robust and flexible, so that a variety of cases could be run without requiring numerous coding changes. A third reason was to keep the analysis simple, allowing us to focus on the dominant physical effects, with minimum distraction from extra factors. To implement this approach we use a number of approximations, which will be described in this section.

The simulation code uses a Smoothed Particle Hydrodynamics (SPH) algorithm for computing pressure forces on a grid with fixed spacing. The adopted size of this spacing was 100 pc, which can be taken as the linear resolution of the models. The code is essentially the same as the code described in detail in Struck (1997), with minor changes described below. The model galaxy is bound by a fixed, rigid dark halo and stellar disc potentials described below. Additionally, gravitational forces are computed between particles in adjacent cells, to capture local gravitational instabilities, though this was not an important aspect of these models.

A total of 437,750 particles were used to represent the disc gas. With the scalings described below the total disc gas mass is $7.2 \times 10^7$ $M_\odot$ in the low-density (LD) case, and 10 times that in the high-density (HD) case. In the low-density case this mass is quite small for a typical disc galaxy in the nearby universe. The gas particle mass was set to a low value in the LD models, which were run first, to minimize the effects of gravitational instabilities within the gas disc, which were not the object of study. These models then provided comparison runs for the high density models run later.

The initial radius of the disc was set to 10 kpc. As can be seen from the figures below, this changed little through the runs. The gas disc was initialized with an empty nucleus (r = 1.0 kpc), a constant gas density core (out to r = 2.0 kpc), and outside that core a 1/r surface density profile. The particles were initialized on a polar grid with a circular velocity appropriate to their initial radius. They were then given a random positional and velocity offset of small amplitude to make their starting distribution less regular. In the outer disc 5000 particles were contained in each annulus of radial width equal to the resolution length of 100 pc. This means that each 100 pc x 100 pc area in the outermost 100 pc annulus of the disc contains about 8 particles, which is rather minimal. However, this latter number scales with 1/r, so the particle resolution is much better at small-to-medium radii. This particle resolution seems adequate for present purposes, but higher particle numbers would be desirable in future models. Overall the disc contains about 31,400 resolution elements in area, so the spatial resolution is good.

## 2.2 Gravitational Potentials and Scale Units

The form of the gravitational potential of the dark matter halo adopted for the model galaxies of the simulations is such that the acceleration of a test particle in halo is,

$$g_h = \frac{GM_h}{\varepsilon^2} \frac{r/\varepsilon}{\left(1 + r^2/\varepsilon^2\right)}, \tag{1}$$

where $M_h$ is a scale mass for the halo, $\varepsilon$ is the core radius (set to 2 kpc), and $G$ is the gravitational constant. The rotation curve v(r) for disc particles bound by this halo is,

$$v = \sqrt{\frac{GM_h}{\varepsilon} \frac{r^2/\varepsilon^2}{\left(1 + r^2/\varepsilon^2\right)}}, \tag{2}$$

which is linearly rising at small radii (r << $\varepsilon$), and flat at large radii ($r$ >> $\varepsilon$). This potential is very similar to the familiar Plummer and logarithmic potentials (e.g., Binney & Tremaine 2008), but the turnover in the rotation curve from linearly rising to flat is more gradual. Since our simulations are confined to within a few disc radii of the potential centre, we do not need a more realistic model of the outer halo. It is assumed that these halo potentials are rigid, in the sense of fixed and unchanging. For simulations of isolated, symmetric discs, there are no global potential perturbations, and this is an excellent approximation. For simulations that use a long distance tidal encounter to drive a persistent spiral wave, the approximation is still a good one, since the perturbation of the inner halo is likely to be small. The barred simulations used a non-axisymmetric halo potential that is described below.

In addition to the dark halo potential, most of our models include a disc component to the total gravitational potential. In most galaxies this potential is largely due to the disc stars, since their mass usually exceeds that of the interstellar gas by nearly an order of magnitude, except in the outer disc, and in low surface brightness galaxies. We do not explicitly model the disc stars, preferring to use our limited particle numbers to better resolve the gas dynamics. Instead we also use a rigid model of the disc potential. In keeping with our approach of making the simulations as simple as possible, we calculate only a vertical (z direction) acceleration of the disc gravity, neglecting sideways (x, y components). We expect this to be a reasonably good approximation for particles that make relatively small excursions above or below the disc (e.g., when $z \ll R = (x^2 + y^2)^{1/2}$). Two obvious cases where this approximation may become inaccurate are when the vertical excursions are large, or generally near the edge of the disc. With regard to the latter case, we are simply not attempting to model the outer gas disc well, which extends beyond the stellar disc.

The disc potential, which accelerates particles in the vertical direction only, has the form,

$$g_d = \frac{-GM_d}{\varepsilon^2} \frac{z/h_d}{|z/h_d|^{n_d}}, \qquad (3)$$

where $M_d$ is the scale mass of the disc, $h_d$ is the scale length of the disc potential, and the exponent $n_d < 1$ determines the form of that potential. In the model of Figure 1 (discussed in Sec. 3.1) we take $n_d = 0.8$. This simple form gives an acceleration that rises rapidly at $z < h_d$ but which increases quite slowly at large $z$, as observed. (See Fig. 1.7 of Spitzer (1978) for classical observations in the solar neighborhood, and Section 10.4 of Binney and Merrifield (1998) for a theoretical discussion.)

Flocculent disc models were run with values of $n_d$ between 0 and 1. The latter corresponds to a constant vertical gravity, e.g., from a thin disc of infinite radial extent. The case $n_d = 0$, corresponds to a linearly increasing vertical acceleration, e.g., from a massive thick disc, or a flattened halo. Realistic profiles have values between these extremes. We experimented with extreme values of this parameter, but in the models described below the value was generally fixed at $n_d = 0.8$. One exception is discussed in Section 3.1, and shown in Fig. 2.

The simulation code is written in dimensionless variables, but all results are presented in physical units using adopted scalings. The adopted scale length is $\varepsilon = 2.0$ kpc. A characteristic (gas thermal) velocity of $c = 6$ km/s was also used, and these imply a time unit of 333 Myr. The mass scale adopted for equations (1) and (2) is $M_h = 2.2 \times 10^{10}$ M$_\odot$. The halo mass contained within a radius of 100 kpc is about $10^{12}$ M$_\odot$. Then the speed of the flat (large r) part of the rotation curve is 220 km/s. In most runs the values of the disc constants were set to $h_d = 400$ pc and $M_d = 1.1 \times 10^9$ M$_\odot$.

**2.3 Cooling and Feedback Heating Terms**

**2.3.1 Cooling**

In keeping with the philosophy above, simple approximations for radiative cooling and young star feedbacks were included in the simulations. These are essential, since without the heating there would be little vertical motion out of the mid-plane. Without cooling, heated gas could not return to a state conducive to star formation. The algorithms used are a compromise between realistic representations of the physics and numerical efficiency. Heating and cooling processes occur over the whole range of scales from atomic to galactic, so inclusion of these processes in all current models of galaxies or galaxy discs is somewhat phenomenological. (Recent illustrations and discussions include: Thacker & Couchman 2000, Springel & Hernquist 2003, Stinson et al. 2006, Saitoh et al. 2008.) A prime example is the fact that the details of how massive stars inject energy and momentum into the ISM via winds and explosions cannot generally be simulated in galaxy scale models at present.

In the simulations the SPH gas particles are assumed to obey an adiabatic equation of state, with internal particle energy added or subtracted according to the heating/cooling functions in a separate step. In the cooling algorithm we approximate optically thin radiative cooling by a simple step function with constant cooling timescales in specified temperature ranges. In each timestep the cooling is implemented by simply multiplying the energy and temperature by $\exp(-\Delta t/\tau_{cool})$, where $\Delta t$ is the timestep and $\tau_{cool}$ is the cooling timescale for the given temperature.

The shortest cooling timescale is applied to particles with temperatures of $4000 < T < 250,000$ K. In this range cooling sources include permitted atomic lines, and at the low end forbidden lines, dust and molecules. The typical timestep of $3.3 \times 10^6$ yr is quite long, except for a very low density or metal-poor gas. Cooling in this temperature range at typical cloud densities would not be time-resolved in our models. Thus, our cooling prescription is a numerical convenience since it allows particle thermal energy loss to occur over tens of timesteps, rather than removing it so rapidly that negative energies result. Since cooling is so rapid in this range, and we do not resolve small scale clumpiness, we do not include a dependence on gas density in the cooling calculation. More details on the cooling algorithm are provided in Sec. 2.3 of Struck 1997.

At temperatures below this range the cooling is reduced by a factor of ten and turned off at a minimum temperature of 1000 K. These values are of little importance to the models, except that low temperatures allow high particle densities. However, SF and feedback heating are triggered at moderate densities, see below. At high temperatures the cooling is set to a value 5 times lower, as a rough representation of the optically thin cooling curve. However, the hottest gas typically cools more quickly via expansion cooling or by doing work on adjacent particles.

**2.3.2 Feedback Prescription and Heating**

Feedback heating is triggered when the internal particle density exceeds a fixed density threshold, and does not have a high temperature (which is almost never the case at high densities). The threshold density in the models presented below is about 21 particles per unit grid cell (or about 0.35 $M_\odot$ pc$^{-2}$ in the LD case and 3.5 $M_\odot$ pc$^{-2}$ in the HD case). This value exceeds the density in most of the initial disc by a factor of a few, except the innermost parts.

This value and the typical surface densities in the disc are very low compared to most late-type galaxy discs in the LD case. It is well known that the limited particle resolution in SPH codes can generate unrealistically strong clumping in some circumstances. See the discussion of Truelove et al. (1997) and the sophisticated treatment in the recent work of Robertson and Kravtsov (2008). By using very low surface densities in the LD models we guarantee compliance with the Truelove et al. criterion (local Jeans length should exceed the grid size) for preventing artificial clumping. However, there are some inconsistencies in this approach, relative to real galaxies. In the LD models the SF threshold has been set at a quite low surface density to generate feedback. Outside the midplane, the space is mostly empty in this case, so there is not much material to resist the motion of gas particles thrown up by this feedback. Thus, the feedback effects are somewhat artificially facilitated in two ways in the LD models.

To return to the problem of artificial clumping, since our code only computes local self-gravity on the grid scale, we greatly reduce its effects (even in the HD models), and so most clumping is initially either a stochastic or wave-driven effect. Once clumps form they do generally have significant self-gravity, until disrupted. The goal in these models was to focus on the wave-driven or large-scale stochastic effects, without having to worry about phenomena generated by artificial clumping.

The cool, dense interstellar cloud complexes where stars form are very clumped, so one might worry that the artificial smoothness of our gas discs might be responsible for unrealistic effects. Those worries, and a desire to have more realistic surface densities and star formation rates, motivated us to run the HD models. The HD models were run with surface densities and threshold densities increased by an order of magnitude, which are much closer to the values of typical discs (see e.g., Kennicutt 1989). These HD models run close to the Truelove limit, and as expected are visibly more clumped. Nonetheless the large-scale characteristics of the EDIG gas and the primary results discussed in the following sections did not change. Comparison of the models lead us to conclude that it is the amount of SF and heating that is important, not the surface density per se.

Feedback heating is applied for a duration of 30 timesteps or a bit less than $10^7$ yr. The choice of this timescale is a compromise, since it is several times the lifetime of an O-type star, but considerably less than the lifetime of the least massive star that can produce a Type II supernova. During the heating period, cooling is turned off for the particle; this assumes that on average heating effects overcome cooling in this environment. Typically about 1% of the particles are in an SF heating state. Dividing the corresponding mass by a mean timestep and assuming a gas-to-star conversion efficiency of a few percent gives a typical SFR of about 0.5 $M_\odot$ per year in the LD models (or about 5.0 $M_\odot$ per yr in the HD models).

The lower right panel of Figure 1 shows the range of temperatures as function of radius with the adopted heating and cooling terms, at one time in one representative model. The first thing that is obvious from the figure is that there is a large range of gas temperatures at all radii. There is more hot gas at smaller radii where the SF is stronger. The amount of million degree gas is unrealistically high (about 8% of the particles), because this particular model also has a relatively high rate of SF. The stepwise cooling curve is reflected in the sharp horizontal discontinuities in the gas density in the lower right panel. Though clearly inaccurate in some ways, our simple cooling/heating terms do produce a continuous range of thermal phases, in a stable configuration, that captures a number of the essential characteristics of the diffuse gas in galaxy discs.

## 2.4 Graphics Details

Here we provide a few additional details on the production of graphs used below for reference. The first of these is simple – all particle plots, including Figs. 1-3, and 5 were made with every tenth particle of the simulation, except for Figure 8, which was made with every fifth particle. Thus, the particle numbers in general, and especially the number of star-forming particles, are much greater than shown in those figures.

Other details concern plots produced by binning gas particles, beginning with the rotation curve panels in the figures listed in the previous paragraph. The division of the particles into separate horizontal layers for the different rotation curves is described in the caption to Figs. 1 and 2. Each layer is divided into concentric annuli of width 0.4 kpc, or 25 annuli within the 10 kpc disc. The particles within each annular layer are identified, and the average of the azimuthal velocities is defined as the azimuthal velocity of that annulus in that layer. Particle numbers tend to be low in the innermost annulus, but high enough to define very good averages in all other annuli.

A similar annular binning procedure was used to produce the mean velocity dispersion plot (Figs. 4, 6) and the mean vertical displacement plot (Fig. 9). In those plots there was no division into parallel horizontal layers, but each annulus was divided into 12 equal angular sectors. Thus, each bin is an angular sector within an annulus, and the bin size increases with radius. Given the initial 1/r particle distribution this means that all bins had roughly equal particle numbers (except in the inner few annuli). With 12 x 25 = 300 bins, we have about 1460 particles per bin, so the random statistical deviations about the means should be small (a few percent). Some idea of the intrinsic dispersions about the mean can be obtained by comparing the curves for adjacent annuli.

## 3 MODEL RESULTS AND ANALYSIS

### 3.1 Flocculent discs

In a first series of models, no waves were imposed on the flocculent initial gas discs. Once star formation (SF) started in the initially quiescent model discs there was a period of transient evolution. However, the discs settled fairly quickly to a global steady state, with small variations around a nearly constant and moderate rate of SF and consequent energy input to the gas. Once this steady state was reached a moderate number of gas particles were displaced substantial distances vertically. The net amount of SF was varied by changing the value of the density threshold for triggering SF in different models. Models with very high rates of SF tend initially to develop transient winds or gusts out of the disc, but not an extensive, quasi-stable layer that we envision the EDIG to be.

The example shown in Figure 1 is from a model with one of the highest rates of SF in this series, and has one of the largest amounts of EDIG ($n_d = 0.8$, see Eq. (3)). It conflicts with observation in several ways. First, the EDIG is centrally concentrated like a wind, though at the late time shown, transients have damped away. Secondly, the EDIG is supported by substantial in-situ SF, even at considerable distances from the mid-plane. This is true to some extent in all the models described below, and the existence of some high altitude SF is observed (Tuellmann, et al. 2003) but this model has a very high amount. Thirdly, as shown by the last panel of Fig. 1, the anomalous rotational EDIG kinematics is not reproduced in this model. The symmetric disc SF displaces gas upward, but as in the ballistic model it does not greatly disturb its rotation.

The dependence of the gravitational potential of the disc on distance from the central plane was also varied. Steeply rising potentials could effectively prevent the gas from traveling very far from the central plane, as expected, and the enhanced compression in the disc did not increase the SF significantly. If the gravitational potential of the disc is sufficiently shallow, high rates of SF drive a wind. Wind material moves far from the disc plane, and so, should not be identified with the usual EDIG. E.g., in the case of no disc gravity, a strong wind developed, which rapidly depleted the disc gas. These results suggest that over a wide range of conditions (including some rather unphysical disc potentials), flocculent disc galaxies will generally not have large amounts of EDIG.

Models with unusually strong potentials, rising steeply with vertical distance from the disc, could only be produced by a massive and thick (stellar) disc. Though unrealistic, given a strong SFR, they can show some very interesting dynamical behaviors, as illustrated in Fig. 2. This figure is produced from a model where the vertical gravitational acceleration rises as the square of the distance from the mid-plane ($n_d = -1$). The SF behavior is cyclic. Rings of strong SF form in the centre and propagate outward, as shown in the second row of Fig. 2. Gas behind the waves is thrown vertically out of the disc to considerable distances considering the strength of the potential.

By the time the ring wave reaches the outer edge of the disc, a large fraction of the gas has been heated and driven away from the mid-plane. Mid-plane gas densities are then much less than the threshold value and SF largely ceases, as in the first and third rows of Fig. 2. The gas cools and falls back to the mid-plane, first in the inner parts, where it has had longer to cool. Thus, SF is reignited in the inner parts, starting another outward propagating ring-wave. In the absence of gas consumption, the cycle repeats indefinitely.

In all our flocculent models either a steady, uniform state or a periodic one like this last, resulted. In the former case, the state was usually characterized by a low SFR, with relatively little EDIG.

Panel C of Fig. 2, like the corresponding panel of Fig. 1, shows that the EDIG rotation curve differs little from the mid-plane rotation curve at all times. The effects of the propagating ring waves on the rotation curve are evident, but they have the same effect on material at all heights. This supports the notion, discussed below, that asymmetric (azimuthally dependent) radial motions of gas elements of the EDIG are an important part of the anomalous kinematics.

We should note one more feature of these models; the EDIG is a generally continuous layer. Figs. 1 and 2 show that it is not concentrated above or near regions of strong SF, as might be expected if the EDIG originated in a small number of strong local fountain flows. The higher surface density models mentioned in Sec. 2.3 above do have a clumpier EDIG, as well as clumpier SF regions in the disc. The recent high resolution H$\alpha$ observations of Rossa, Dahlem, Dettmar, & van der Marel (2008) also find an intricate filamentary network in the EDIG. The structure in the HD models and the new observations is similar and deserves further investigation. The main point here is that neither the models nor the observations show discrete clumps or streams of EDIG, but rather a continuous layer.

A related question is – what is the probability that a gas particle in or near a star-forming region will be kicked upward into the EDIG? Hydrodynamic models of individual disc fountains or superbubble break-out show that this probability is about 50% (see the recent review of de Gouveia Dal Pino et al., 2008, and references therein). In these models the fraction is determined largely by how much energy and momentum are expended in clearing a blow-out path through assumed hot overlying layers. After the EDIG layer has been established, similar phenomena occur within our models. We do not have the particle resolution to properly resolve local fountain flows well. However, our gas, which is clumpy both within and around the disc, is in some ways more realistic than the smooth, stratified medium used in most studies of individual blowouts. Nonetheless, by studying particle trajectories (like that in Fig. 7), we find that the probability of kicking a particle significantly away from the mid-plane is also roughly 50% in our models. Thus, although the processes may be different the effect is of a similar magnitude in both types of model. Future models with a factor of a few better spatial resolution and a more careful treatment of the feedback processes could bridge the gap between these two types of simulation, but probably won't yield a very different result.

**3.2 Driven spirals**

We carried out two series of simulations with waves imposed on the gas disc. Both types of model provide evidence for the conclusion that a vigorous EDIG is a wave driven phenomenon. In this section we describe models in which spiral density waves are driven by the tidal force of a companion galaxy of modest mass. The disc of this model was initialized like those of the previous subsection, except that an orbiting companion halo potential was added. This halo companion was placed on a circular orbit, at a distance of 40 kpc from the centre of the disc on the x-axis. Its mass was about half that of the primary and the form of its gravitational potential was the same as that of Eq. (2). Because of its large distance from the disc centre it moved relatively little over the course of the simulation, and so, can be viewed as a nearly static perturber, which induces spiral density waves in the disc. This is an easy way to induce such waves without a fully self-consistent, self-gravitating simulation. After an initial phase of transient relaxation, the spiral waves develop and persist as relatively stable features throughout the run.

Figure 3 shows the HD model disc at a couple of relatively late times, when the wave is well developed and the SF is relatively strong. The SF and EDIG measures are observed to settle to fairly constant values after initial transients damp. Despite the strong star formation, and a weaker vertical potential for this model, the gas is not thrown as far out of the mid-plane as in the model of Fig. 2. The SF in the present model is not as symmetrical as in that model, and evidently gas particles do not receive an equally strong boost out of the central plane. On the other hand, the two C panels of Fig. 3 show that the EDIG in this model is rotating significantly slower than that in the mid-plane (with values intermediate between the flocculent and the barred example discussed below). Figure 4 shows how the SF concentrated in the wave generates turbulence in the gas, as measured by the particle velocity dispersion, which peaks in the wave region over a large range of radii. (An LD model is shown in Fig. 4; the corresponding HD model yields a very similar picture.)

Although not shown in Fig. 4 we have also compared an in-plane component of the velocity dispersion (the x-component) to the vertical component of the velocity dispersion. The x-dispersion is slightly greater (a few km s$^{-1}$) than the z-dispersion at most radii. This statement is also true for flocculent models. The amplitude of the x-dispersion is highest in the waves. This statement is not true for the flocculent models, where the two dispersions are similar around most annuli. The wave perturbation in the disc plane seems to be the dominant factor in causing this difference.

A real case of an edge-on system (NGC 5529) involved in a galaxy interaction, with possible wave driving as in the present model, has been described by Irwin et al. (2007). EDIG is evidenced by the presence of hydrocarbon molecules some kiloparsecs above the disc. Recently, Mapelli, Moore & Bland-Hawthorne (2008) have suggested that the lopsidedness of the prototype EDIG galaxy NGC 891 is due to a flyby interaction. The models suggest that more observations of edge-on, interacting systems are warranted, with the caution that central starbursts, which are often triggered by strong interactions, drive winds, which will complicate the interpretation.

**3.3 In a barred potential**

In this section we discuss a model in which the background halo or stellar potential is not axially symmetric, but has a bar-like form. This is implemented by introducing asymmetries into the halo potential of Eq. (1). Specifically, we used the form,

$$g_h = GM_h \frac{r/\varepsilon}{\left(\varepsilon^2 + 1.5x^2 + 0.667y^2 + z^2\right)}, \qquad (4)$$

where the coefficients of the x and y terms (1.5 and 0.667) are arbitrary, except for the requirement that they differ from 1.0 sufficiently to give an easily visible effect. All the other constants are the same as in Eq. (1). The vertical disc potential of Eq. (3) was not changed in this model. It is not correct to maintain a cylindrically symmetric disc potential when the shape of the disc is affected as much as shown in Figure 5. However, the main effects would be in underestimating the restoring force in the dense bar region (an effect which could be naturally offset by additional feedback), and in overestimating the restoring force in regions away from the dense bar.

The initial positions and velocities of the gas particles are assigned to near circular orbits not in equilibrium with the bar potential. There is a period of relaxation for the disc to settle into the bar-like configuration. The initial circular orbits quickly evolve to more elliptical orbits, modulated by shocks in large-scale waves. Thus, the disc assumes an elliptical form along equipotential contours, but with a bar-like wave, which propagates through the oval disc (see Fig. 5).

In Fig. 5 the panel A series shows the rotation of the disc bar through more than 150°, from the first to last row. The panel B series shows only moderate variations from timestep to timestep. That is, the structure and extent of the EDIG does not vary greatly, where it is present. However, in the first and last timesteps there is little EDIG at large values of |x|. Comparison to the corresponding A panels shows that at those times the projection of panel B is looking down the bar, and thus, there is evidently little EDIG outside the region of the bar.

The SF is concentrated in this wave, and the EDIG is strongly concentrated in the vicinity of the wave. Therefore, the EDIG covers only part of the gas disc at any time, settles back to the central plane in regions away from the bar, and is continuously regenerated on a wave propagation timescale.

The first row of Figure 5 shows that the rotation speed of the EDIG can be considerably reduced relative to the central plane value, e.g., by an amount comparable to the observed few tens of km s$^{-1}$ (see e.g., Heald et al. 2006 (their Fig. 2), Oosterloo, Fraternali & Sancisi 2007 (their Fig. 15)). Figure 6 shows a similar velocity dispersion profile to Fig. 4, with turbulence generated in the wave in both cases. It appears that the reduced rotation is the result of a wave-induced sequence of events: wave compression triggers SF, energetic young star feedback generates turbulence, the turbulent pressure pushes gas out of the central plane, and the turbulent pressure gradient modifies particle orbits. Most importantly, the asymmetric distribution of these forces allows for perturbations of the orbits, with the resulting anomalous kinematics.

To return to velocity dispersions for a moment, a comparison of the in-plane component of the velocity dispersion (the x-component) to the vertical component, as discussed in the previous subsection, yields somewhat similar results. However, the two dispersions are more similar to each other than in the spiral wave case discussed there. Thus, this case is more like that of the flocculent disc. The amplitude of the variation in the dispersion of the two components around an annulus is also quite similar, with the x-component only slightly larger, in contrast to the spiral case where the x-dispersion is significantly larger. One significant difference between the two components in this case is that the z-dispersion is very small at large radii where the SF is small, while the x-dispersion is about the same size at all radii. This suggests that the x-dispersion is maintained by more than just the SF feedback, which is reduced at large radii in this model. We suspect that particles at the outer edge of the disc do not have their non-circular motions damped by as many collisions.

Examples of individual particle trajectories, like that shown in Figure 7, confirm this picture. Generally, before or at the beginning of an excursion out of the plane, a particle experiences collisions and turbulent interactions. These are evident as irregularities in the generally smooth orbit, best shown in Fig. 7 as jitter in the azimuthal velocity. The radius of the particle is not correlated with the onset of these excursions, which is what we would expect for particles whose orbits are precessing ellipses, which intersect the waves at arbitrary phase. However, after a particle reaches its maximum height, with a reduced rotational velocity, and is no longer supported by turbulence (generated by SF feedback in the wave), it generally falls inward in radius, as well as in height (see Fig. 7). There are wide variations in the character of particle orbits during a vertical excursion, but the example in Figure 7 is typical, and helps understand the average behavior.

The panel C series of Fig. 5 shows that there is much variation in the EDIG kinematic anomaly as a function of time, or more accurately, as a function of the orientation of the disc bar. In the first and last two rows of the series the anomaly is substantial. In the middle two rows it is minimal. The corresponding A panels show that the latter case of little anomaly occurs when the disc bar is nearly perpendicular to the halo bar and the long axis of the disc. Specifically, when the bar is aligned with the long axis of the disc, then the SF, the mass of the EDIG, and its extent are maximal (though the first two quantities don't vary by a large amount with time). At such times the rotation reduction is large. When the bar is orthogonal to the long axis of the disc it is short, and the mean rotation reduction is much reduced.

How the bar wave drives the gas is further illustrated by Figures 8 and 9. Fig. 8, shows gas particles with either positive, upward velocities (panel A), or negative velocities (panel B), contained within cylinders whose inner and outer radii are given in the figure caption. The azimuthal position is given on the x-axis, the vertical position on the y-axis. Red symbols denote the hotter particles, and blue asterisks the star-forming particles. These cylindrical slices and Fig. 5 make clear that SF is concentrated at the back end of the bar wave, where particles are flowing into the wave. The SF heats the gas and throws it upward out of the disc with positive velocity (and downward with negative velocity). This gas is visible as the red plumes in Fig. 8. Further downstream the gas begins to cool, but continues its motion away from the disc. This is seen as the increasing fraction of cool particles at positive values of z, located at the back end of the plumes.

Later the cool gas begins to fall back into the disc, which is illustrated by the many cool gas particles located at negative values of z in panel A (at positive z in panel B). There are also a number of hot particles located at negative z values (in panel A). These tend to be found near the SF regions. It seems likely that these are infalling particles that are shock heated when they collide with particles being pushed out by the SF.

This picture of the effects of the waves is confirmed by Fig. 9, which shows the mean vertical displacements of the gas particles (with positive z velocities) at different azimuths on curves of constant radii. Successive curves are displaced vertically, so dashed lines are used to show zero displacement for each curve. The inner and outermost few curves are somewhat anomalous, but the middle curves are more representative. Like

Fig. 8, they show mean outflow displacements of the particles in the waves, i.e., at azimuths a bit less than 90° and 270° (also see Fig. 5). Also in accord with Fig. 8, these curves show mostly negative displacements (inflow) outside of the wave. The magnitude of these displacements is about 1.0 - 2.0 kpc from the zero displacement line, in accord with the observed scale of the EDIG.

Finally, we note that the number of star-forming particles in this model is roughly constant with time, after early transients have damped away. Given the changing length of the bar evident in Fig. 5 this may seem surprising. On the other hand, it is also apparent in Fig. 5 that the SF particles are denser when the bar is shorter. (However, the model was not run long enough to allow us to exclude a small amplitude mean variation with bar phase.) The near constancy of SF is a feature that this model shares with all the other models – isolated discs and discs with tidal spirals. The mean level of SF varies between models, and with the values of the SF feedback parameters. However, the general tendency is for self-regulation to a mean value. The only clear exception is when the disc potential and feedback parameters are such that large quantities of gas can be thrown to large scale heights, so that the SF is characterized by bursts separated by long quiescent periods (as discussed above).

## 4 SUMMARY AND CONCLUSIONS

The picture suggested by the models above is that the EDIG and its rotation reduction are wave driven phenomena, whose strength correlates with that of the driving waves. We caution that models above are meant to exploratory and are very simple in a couple of areas. These include the rather schematic heating and cooling algorithms, and the absence of full self-gravity. We do not think that improved models will change the basic results above, but they will be needed for comparisons to increasingly detailed observations of the EDIG.

There are a number of consequences of the proposed correlation between disc waves and EDIG, including the following.

**1.** The energy needed to support the EDIG layer is considerably less than if this layer covered the whole disc. For example, in Fig. 5, if one considered a circular disc with a radius of about 10 units, then at most time the area dominated by green particles (the EDIG) in the figure would be roughly half the disc. Alternately, to get a similar column density of EDIG (seen edge-on) from a cylindrically symmetric, flocculent disc would probably require a lower mean surface density of SF than in the waves of Fig. 5. However, it would likely require a somewhat higher net amount of SF in order to propel similar amounts of gas to similar heights over a larger area, since the expulsion of any gas element would not benefit from synergistic inputs from the more closely packed heat sources in the waves.

The results in spiral wave model of Sec. 3.2 are quite similar to those of the barred models described in Sec. 3.3. This despite the fact that the details of the driving force are different, e.g., in the former model the primary halo potential is not perturbed. We conclude that the EDIG kinematic anomaly is the result of the wave in both cases, not the details of the driving mechanism.

**2.** The EDIG is not a static structure. It is steady, but locally is as dynamic as the waves that drive it. Its structure is analogous to the tops of storm clouds along a frontal boundary in terrestrial weather (and much as envisioned by Norman and Ikeuchi (1989)). This conclusion, and the idea of wave-driven EDIG, could be tested by comparing the fraction of edge-on galaxies with strong EDIG to the fraction of face-on galaxies with strong bars or spiral waves. Evidence already exists for the presence of EDIG in a number of partially inclined galaxies with bar components (Fraternali, Oosterloo, Binney, & Sancisi 2007, Boomsma et al. 2005). Very recent survey results suggest that HI clouds in the lower halo of our Galaxy are linked to spiral features (Ford et al. 2008).

**3.** This correlation provides a clear mechanism for global self-regulation in star-forming galaxy discs. In the present universe, much of the surface area of discs in normal late-type galaxies is below threshold densities for vigorous SF, so self-regulation is inactive. However, one can consider a gedanken experiment in which the gas density is suddenly increased by a large amount above threshold in a patch on the disc. Vigorous SF would be triggered, EDIG would form above the patch and spread the gas over adjacent regions, ultimately smoothing away the over-density and regulating the SF. (Further discussions of global turbulent self-regulation can be found in Silk (1997), Struck & Smith (1999) and Koyama & Ostriker (2008). The operation of this process is also implicit in many recent gas disc evolutionary models.)

**4.** Conditions in the EDIG are not conducive to SF, so the EDIG is essentially fuel storage for a limited time. There is a connection to galactic winds, which propel entrained cool gas to higher altitudes where the return time is longer. Winds also contribute to the extended hot halo, where the cooling time is much longer than in the EDIG (Putnam, Grcevich, & Peek 2008).

**5.** The models suggest that large-scale, three-dimensional motions are an essential, not peripheral, aspect of gas dynamics in galaxy discs.

**6.** We would expect self-regulation to be much more active in young, gas-rich discs. Such discs would likely be characterized by high SF rates (see e.g., Elmegreen et al. 2007), which would generate massive EDIG layers. The EDIG is much less conducive to SF than cold clouds in the central plane, so its production is equivalent to putting part of the fuel for SF into storage, and tempering the disc SF. Both the EDIG flows and the turbulence that generates them work to smooth and distribute the gas, as per the above example. Some SF may occur in EDIG layers (Tuellmann et al. 2003), and in the models as shown in Figs. 1, 3 and 5. The strength of this process in young discs may partially account for the formation of the thick component of stellar discs. Many questions remain and we believe that the results above open many avenues for additional research.

**Acknowledgements**

We acknowledge support from NASA Spitzer grant 1301516, and helpful comments from Beverly Smith, William Gutowski, Charles Kerton and Lee Anne Willson. We are also grateful to the referee for suggestions that considerably clarified the presentation.

**Figures**

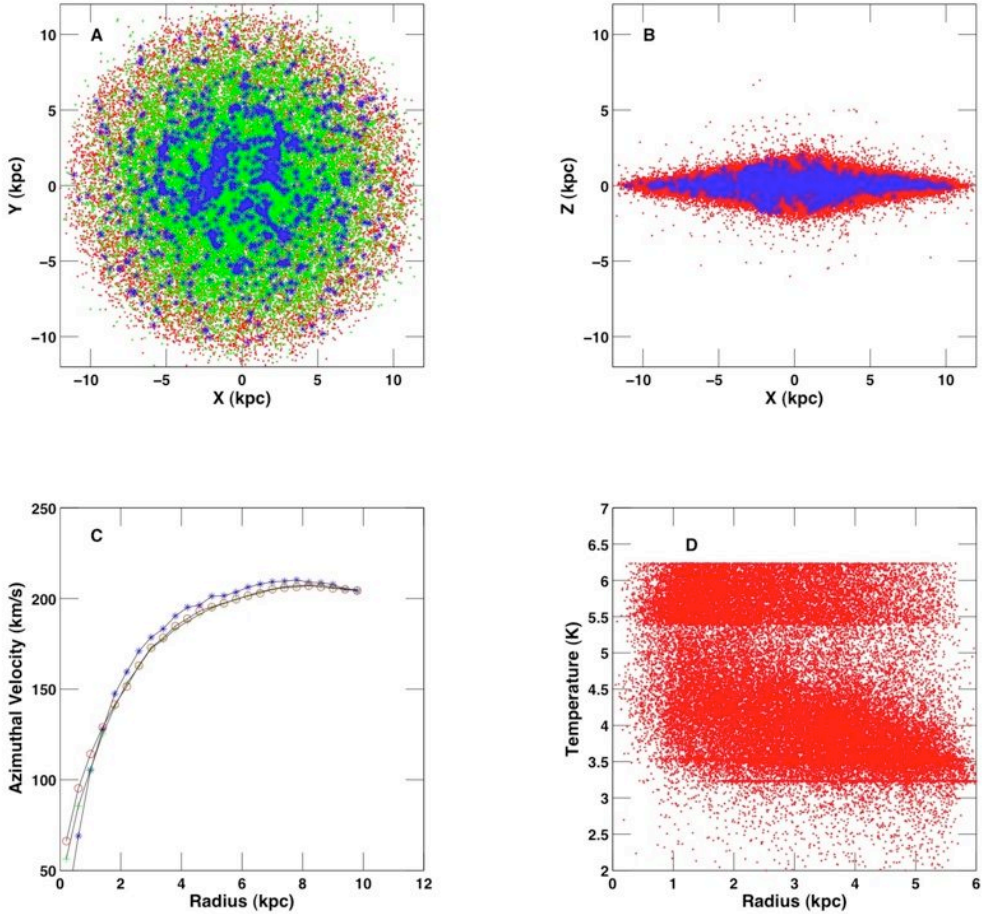

**1.** Three views of an LD (low density) SPH model disc (mean surface density of 0.35 $M_\odot$ pc$^{-2}$). This model has strong, flocculent SF (e.g., at the time shown the SFR is about three times the LD mean of 0.5 $M_\odot$ per year), and is shown after transients have relaxed. In (A) and (B) every tenth gas particle of the model is plotted for views onto the disc plane, and perpendicular to it, respectively. Blue asterisks denote sites of active SF (and energy input). In (A) green points represent particles at a distance of 600 pc above or below the disc mid-plane (i.e., much of the EDIG). Red points represent particles close to the plane in (A), and all non-star-forming particles in (B). Note the general symmetry of SF particles in (A) and their vertical displacement in (B). Panel (C) gives rotation curves in three horizontal slices at different vertical heights: within 300 pc of the mid-plane (asterisks), at 300-900 pc above or below the mid-plane (plus signs), and at 900-1500 pc above or below the mid-plane (circles). Panel D shows the distribution of gas particle temperatures as a function of radius at the same time. See text for further details.

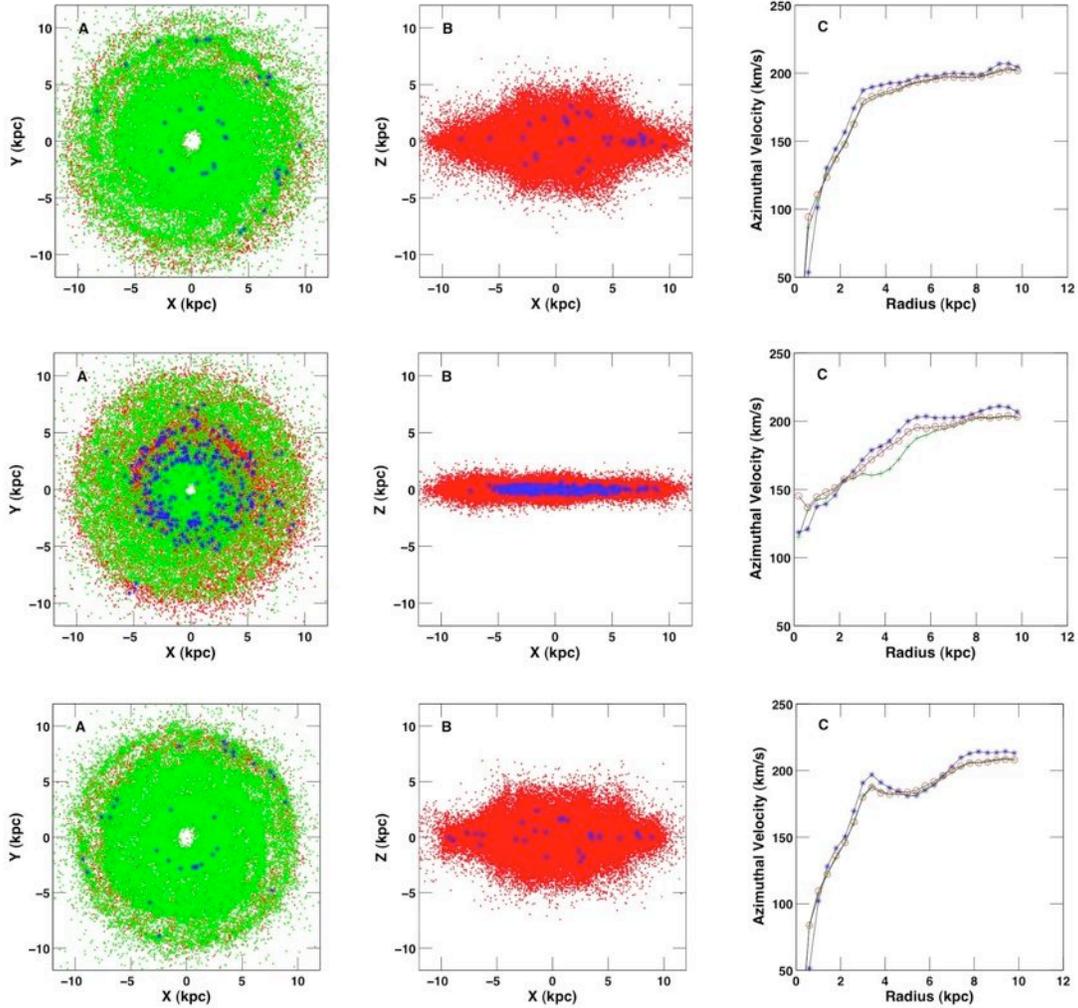

**2.** Same as Fig. 1, but for a model with a steeply rising potential perpendicular to the disc. In each panel (A) and (B) every tenth gas particle of the model is plotted for views onto the disc plane, and perpendicular to it, respectively. Blue asterisks denote sites of active SF (and energy input). In panels (A) green points represent particles at a distance of 600 pc above or below the disc mid-plane. Red points represent particles close to the plane in (A), and all non-star-forming particles in (B). Panel (C) gives rotation curves in three horizontal slices at different vertical heights: within 300 pc of the mid-plane (asterisks), at 300-900 pc above or below the mid-plane (plus signs), and at 900-1500 pc above or below the mid-plane (circles). The timesteps shown are 297, 327, and 353 Myr from the start of the model.

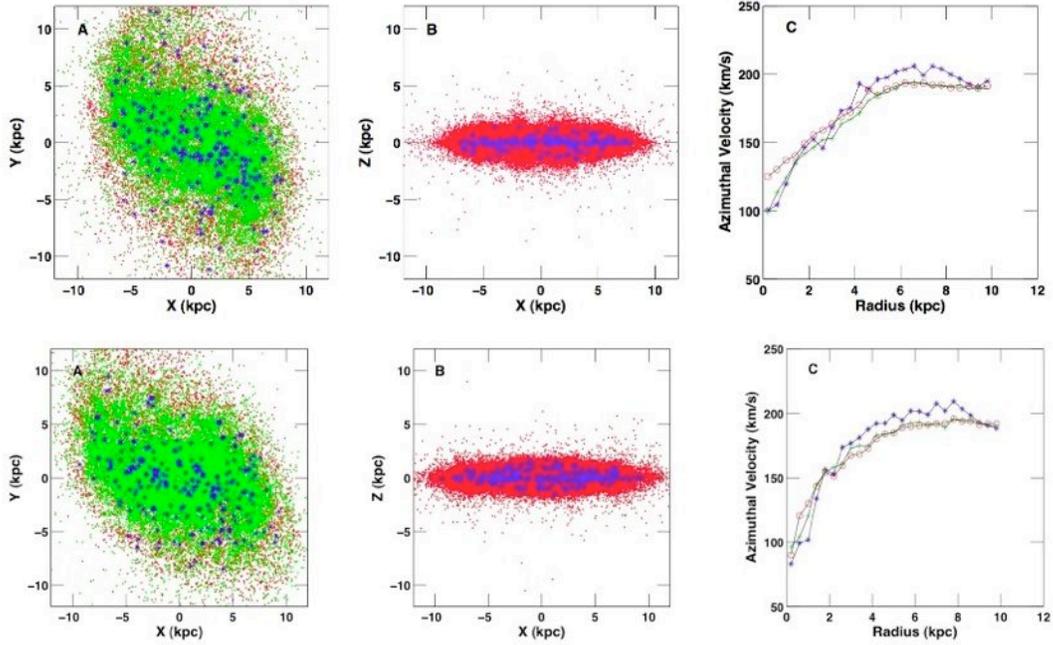

**3.** Same as Fig. 1, except for a (HD, high density case with a mean density of 3.5 $M_\odot$ pc$^{-2}$ and mean SFR of 5.0 $M_\odot$ per year) model with a tidally driven spiral density wave. The timesteps shown are 619, and 667 Myr from the start of the model.

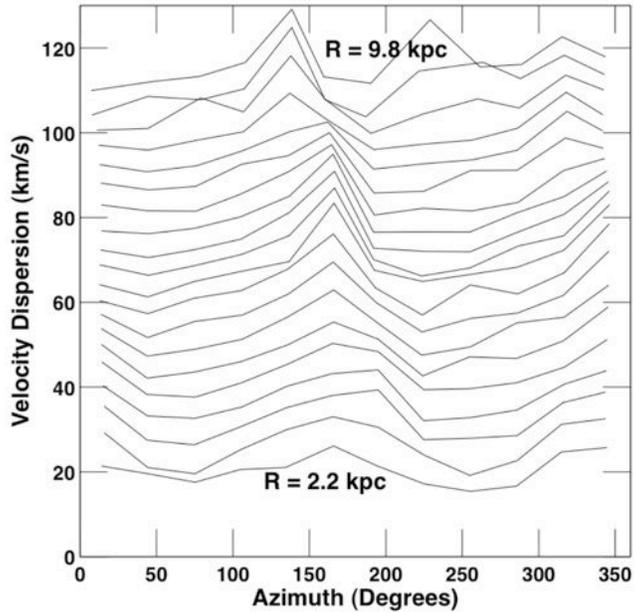

**4.** The three-dimensional velocity dispersion of the gas particles as a function of azimuth in annuli at different radii in the disc, in the driven (low density) spiral model. The mean radius of the annulus represented by the bottom curve is 2.2 kpc. The radial width of the annuli is 400 pc, so the mean radius of each successive curve above the bottom is increased by another 400 pc to a maximum of 9.8 kpc for the topmost curve. Each successive curve from the bottom is vertically offset by 5.0 km s$^{-1}$ to enhance visibility. At all but the largest and smallest radii, the velocity dispersion peaks in the spiral wave.

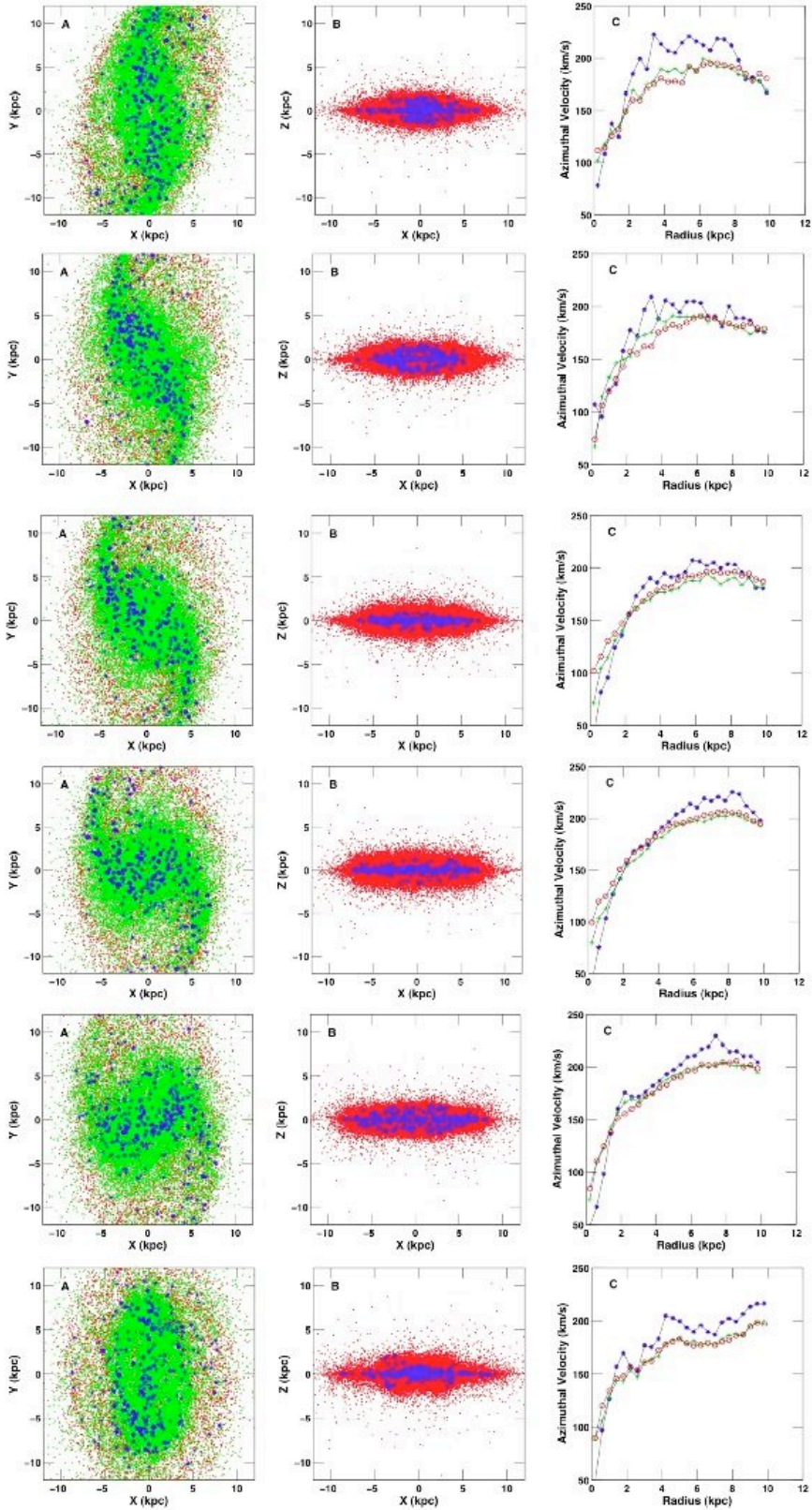

**5.** Same as Fig. 1, but for a (high density disc) model with a rigid bar potential. Symbol definitions for each panel are also the same as in Figure 1. The timesteps shown are 209, 294, 338, 408, 453, and 657 Myr from the start of the model, showing the bar rotation cycle. In (A) panels note the ridges of SF particles (asterisks) marking the bar wave fronts, and the concentration of extraplanar gas in the bar. (B) panels show that the radial range of extensive EDIG coincides with that of the strong SF fronts in (A). Panels C show the rotation reduction of the EDIG layers (plusses and circles) relative to the mid-plane region (asterisks).

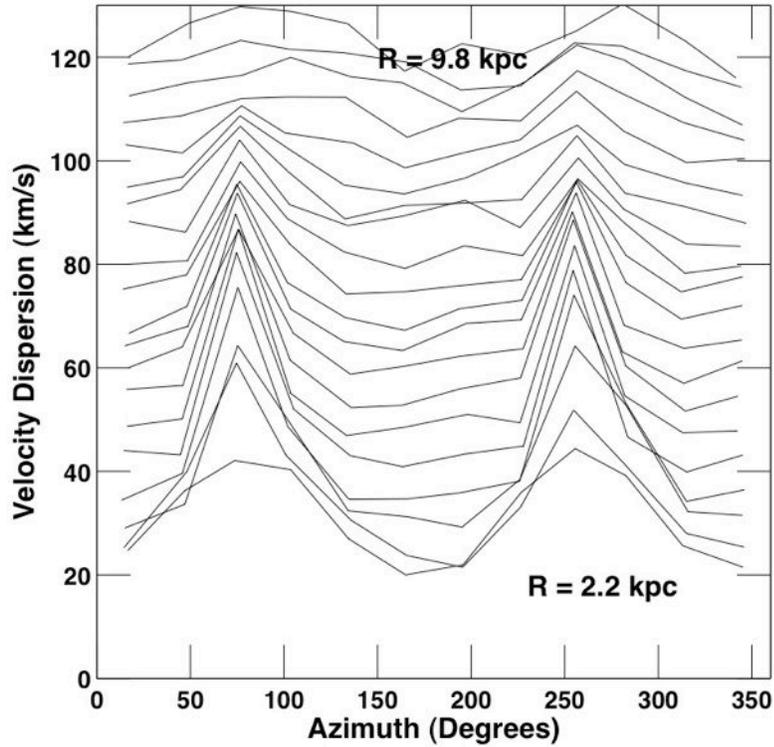

**6.** The three-dimensional velocity dispersion of the gas particles as a function of azimuth in annuli at different radii in the disc at time 610 Myr (as in Fig. 4 but for the high density barred model). The mean radius of the annulus represented by the bottom curve is 2.2 kpc. The radial width of the annuli is 400 pc, so the mean radius of each successive curve above the bottom is increased by another 400 pc to a maximum of 9.8 kpc for the topmost curve. Each successive curve from the bottom is vertically offset by 5.0 km s$^{-1}$ to enhance visibility. At all radii the velocity dispersion peaks in the bar wave.

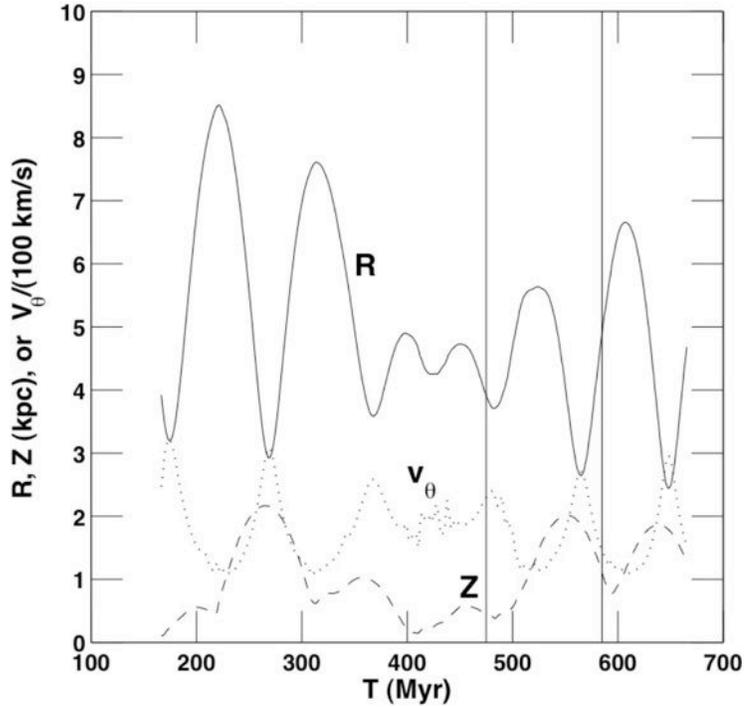

**7.** The variables describing the trajectory of a single particle as a function of time in the (high density) barred model. The solid curve shows the disc radius R (i.e., the radius when the trajectory is projected onto the disc). The dashed curve shows the absolute value of the vertical coordinate z. The dotted curve shows the magnitude of the azimuthal velocity, $v_\theta$ (in units of 100 km s$^{-1}$). Early times (transient phase) omitted. The region between the vertical line illustrates a typical particle excursion above the disc mid-plane, which is often preceded by a period of 'jitter' in the azimuthal velocity.

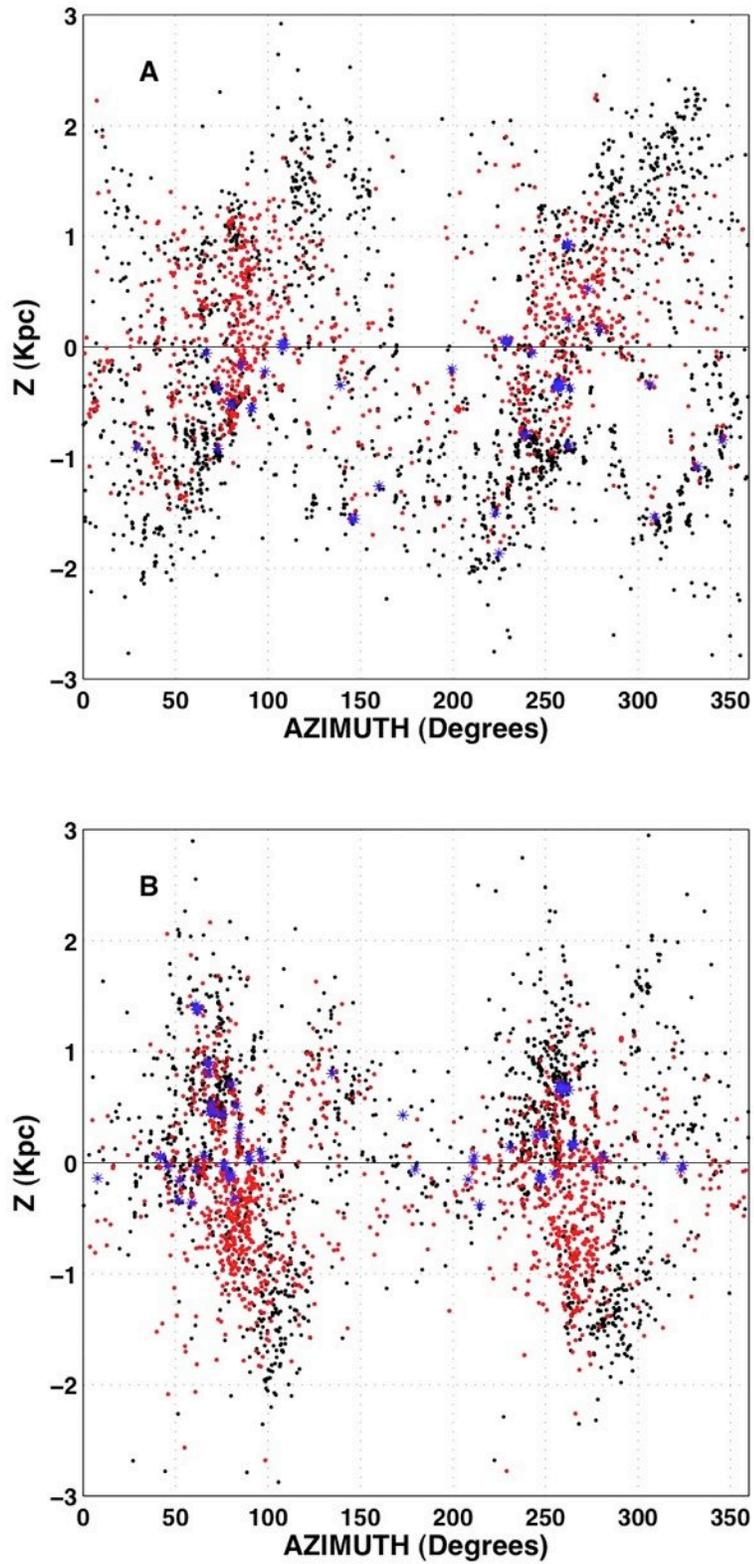

**8.** Particles (every fifth particle) in two annular sections at a time 610 Myr after the start, from the (high density) barred model. In both cases red symbols denote hot particles with more than five times the initial gas particle thermal energy or temperature. Black symbols denote the cooler particles. Blue asterisks denote particles in which star formation was recently initiated. The top (A) panel contains particles located between mid-plane radii of 3.5 and 4.5 kpc; the bottom panel (B) contains particles located between mid-plane radii of 5.0 and 6.0 kpc. In panel A only particles with positive, upward vertical (z) velocities are plotted. To show the symmetry, in panel B only particles with negative vertical velocities are plotted. See text.

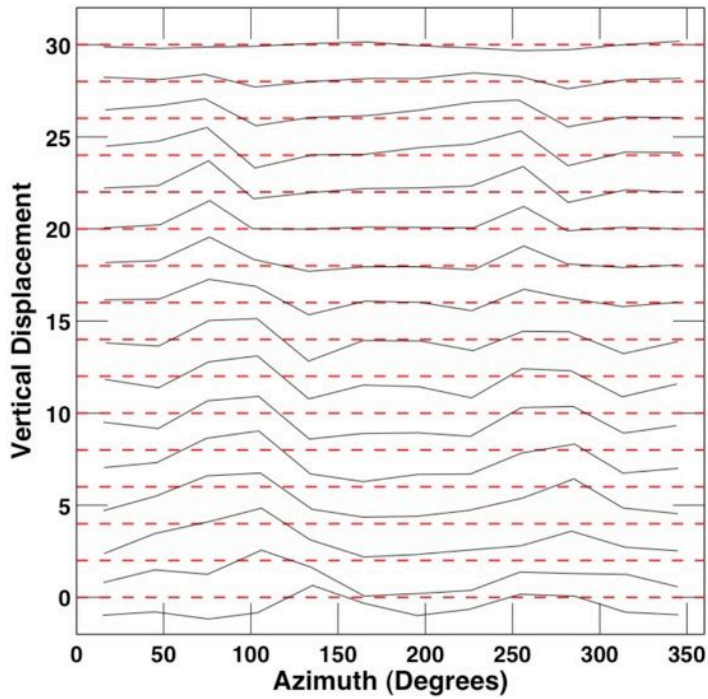

**9.** Mean vertical displacements (in the low-density barred model) of gas particles at time 595 Myr that have been binned on a cylindrical grid consisting of 12 equally sized azimuthal sectors, and with radial bins of width 0.5 kpc ranging over radii of 2 to 10 kpc. Each curve consists of one radial annulus, e.g., the bottom curve represents particles between radii of 2.0-2.5 kpc, and the top particles are at radii of 9.5-10 kpc. Only particles with positive, upward velocities have been included in the averages. Each successive curve is displaced upward by 2.0 kpc for clarity. For reference, dashed, horizontal lines show zero displacement for each curve. All the vertical displacements (in kpc) have been multiplied by a factor of 3.0 to enhance their visibility.